\documentclass[aps,prd,
              reprint,
              amsmath,amssymb,amsfonts,groupedaddress,
              longbibliography,
              superscriptaddress,floatfix]{revtex4-1}

\usepackage{graphicx}
\usepackage{dcolumn}
\usepackage{bm}
\usepackage{amsbsy}
\usepackage{units}

\usepackage{color}

\usepackage[breaklinks=true]{hyperref}
\usepackage{hyperref}
  \hypersetup{
  colorlinks=true,
  linkcolor=blue,
  citecolor=green,
  filecolor=black,
  menucolor=black,
  urlcolor=blue
  }

\newcommand{\sdist}{\kern 0.20em}
\renewcommand{\eqref}[1]{Eq.\sdist(\ref{#1})}
\newcommand{\figref}[1]{Fig.\sdist\ref{#1}}

\allowdisplaybreaks

\usepackage{upgreek}

\begin{document}

\title{Observing Light-by-Light Scattering in Vacuum with an Asymmetric Photon Collider}

\author{Maitreyi Sangal}
\affiliation{Max-Planck-Institut f\"ur Kernphysik, Saupfercheckweg 1, D-69117 Heidelberg, Germany}
\author{Christoph H. Keitel}
\affiliation{Max-Planck-Institut f\"ur Kernphysik, Saupfercheckweg 1, D-69117 Heidelberg, Germany}
\author{Matteo Tamburini}\email{matteo.tamburini@mpi-hd.mpg.de}
\affiliation{Max-Planck-Institut f\"ur Kernphysik, Saupfercheckweg 1, D-69117 Heidelberg, Germany}

\date{\today}

\begin{abstract}
The elastic scattering of two real photons in vacuum is one of the most elusive of the fundamentally new processes predicted by quantum electrodynamics. This explains why, although it was first predicted more than eighty years ago, it has so far remained undetected. Here we show that in present-day facilities, the elastic scattering of two real photons can become detectable far off axis in an asymmetric photon-photon collider setup. This may be obtained within one day of operation time by colliding 1~mJ extreme ultraviolet pulses with the broadband gamma-ray radiation generated in nonlinear Compton scattering of ultrarelativistic electron beams with terawatt-class optical laser pulses operating at a 10~Hz repetition rate. In addition to the investigation of elastic photon-photon scattering, this technique allows us to unveil or constrain new physics that could arise from the coupling of photons to yet undetected particles, therefore opening new avenues for searches of physics beyond the standard model.
\end{abstract}

\maketitle

In classical electrodynamics, light beams in vacuum pass through each other unaffected~\cite{landau-lifshitz2}. However, in the realm of QED photons in vacuum can scatter elastically via virtual electron-positron pairs~\cite{Berestetskii-book}. This qualitatively new process has profound implications in cosmology, where it results in a modified photon spectrum~\cite{zdziarskiNPB89, svenssonAJ90}, as well as in astrophysics, where photon-photon scattering can occur in the strongly polarized vacuum present in the magnetospheres of pulsars and magnetars~\cite{ritusJSLR85, marklundRMP06, dipiazzaRMP12, dunneEPJST14}. 

Notably, photon-photon scattering also plays a critical role in searches of physics beyond the standard model. In fact, photon-photon scattering could also be mediated by unknown weakly interacting particles such as axionlike particles, minicharged particles, magnetic monopoles or hidden gauge bosons~\cite{giesPRL06, abelPLB08, tommasinJHEP09, jaeckelARNPS10, villalbaJHEP13, villalbaPLB16, beyerPRD20} (see also the recent reviews~\cite{MarshPR16, ferreiraXXX20}). These particles occur in many theories of physics beyond the standard model, and are thought to constitute some or all of the dark matter of the Universe~\cite{ringwaldPDU2012}. Indeed, despite their very weak coupling, these exotic hypothetical particles could lead to observable astrophysical and cosmological effects such as matter-antimatter asymmetry~\cite{coPRL20} or the recent evidence of birefringence in cosmic microwave background radiation~\cite{minamiPRL20, fujitaXXX20, takahashiXXX20}. Moreover, such beyond-standard-model particles can be light and cold, which renders them difficult to detect with current techniques based on nucleon and electron scattering because of their small recoil~\cite{essigP20}. Yet, if these particles exist, they provide anomalous contributions to rare vacuum processes such as elastic photon-photon scattering, therefore altering the number and distribution of scattering events predicted by QED.

Following the seminal calculations of photon-photon scattering by Euler~\cite{eulerAP36} and the full QED results by Karplus and Neuman~\cite{karplusPR50, karplusPR51, detollisNC65, Berestetskii-book}, proposals to detect the elastic photon-photon scattering of synchrotron radiation~\cite{csonkaPRD74} or of free electron laser (FEL) radiation~\cite{beckerPRA88, beckerJOSAB89} were suggested. More recently, the advent of high-power laser systems has led to ingenious proposals to detect the finite photon-photon coupling induced by the presence of virtual particles by observing four-wave mixing in the collision of three intense laser beams in vacuum, which results in signal photons whose frequency and propagation direction differ from those of the colliding beams~\cite{moulinOC99, lundstromPRL06, lundinPRA06, kingNP10, kingNJP12, giesPRD18, blinnePRD19}. Other proposals suggested to employ cavities~\cite{brodinPRL01, erikssonPRA04}, Bragg interference~\cite{kryuchkyanPRL11}, or to measure phase correction~\cite{tommasiniPRA08}, vacuum polarization and magnetization~\cite{mondenPRA12}, as well as vacuum birefringence and dichroism induced by ultraintense laser fields~\cite{dipiazzaPRL06, dinuPRD14, mohammadiPRA14, karbsteinPRD15, kingPRA16, nakamiyaPRD17, braginPRL17, mosmanPRD21}.
Remarkably, the detection of elastic scattering of quasireal photons in ultraperipheral Pb nuclei collision was proposed~\cite{denterriaPRL13} and then discovered in the Large Hadron Collider~\cite{aaboudNP17, aadPRL19, sirunyanPLB19}, where an 8.2 standard deviation significance and $Q^2< 10^{-3} \text{ GeV}^2$ virtuality were attained~\cite{aadPRL19}. However, this important result still needs to be independently verified in direct measurements of elastic \emph{real} photon scattering. Indeed, the interaction of real photons in vacuum has never been observed directly. To date, x-ray beam experiments could only place an upper bound on the photon-photon scattering cross section~\cite{bernard2000, inadaPLB14, yamajiPLB16, inadaAS17}.

\begin{figure*}[tb]
\centering
\includegraphics[width=\linewidth]{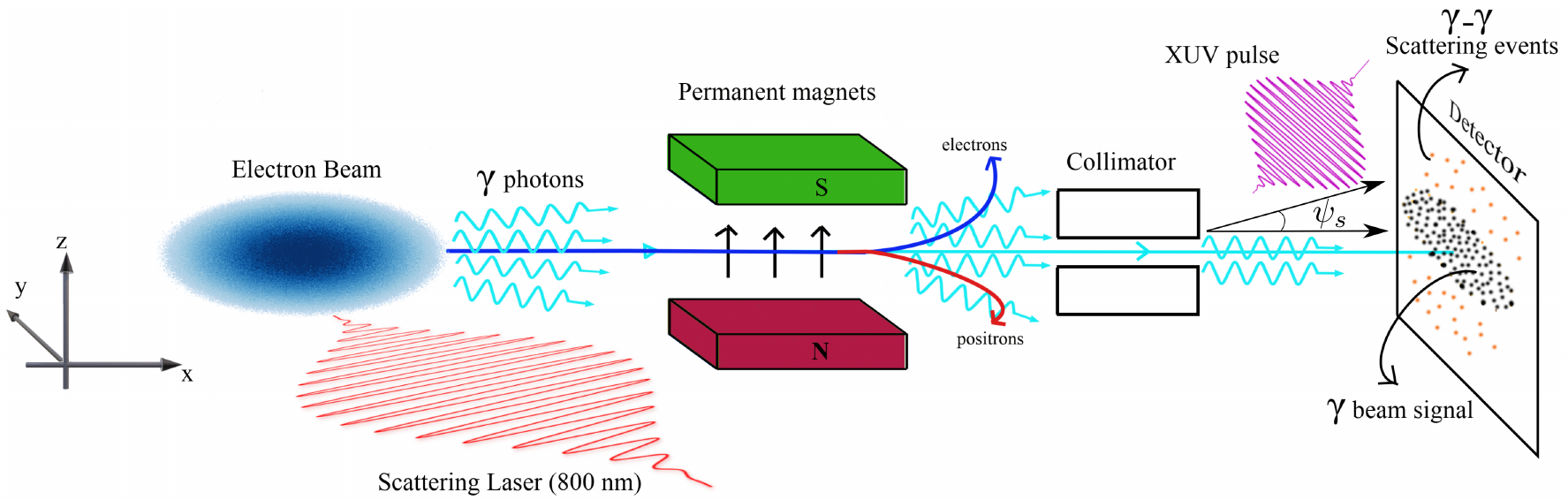}
\caption{Schematic setup. A collimated ultrarelativistic electron beam propagating along $x$ collides with an intense linearly polarized optical laser pulse therefore generating a collimated gamma-ray beam. The laser pulse polarization axis and propagation direction are in the $xy$ plane. The electron beam and possible positrons generated in the collision with the laser pulse are deflected by a magnetic field along $z$. The gamma-ray beam passes through a collimator, which selects photons with small angle with respect to the $xy$ plane. The collimated gamma-ray beam collides with an XUV pulse propagating in the $xz$ plane. Elastic photon-photon scattering events can result in large-angle $\psi_s := \arctan{(p_{s,z}/p_{s,x})}$ photon deflection, where $\mathbf{p}_s$ is the photon momentum after scattering.}
\label{fig:1}
\end{figure*}

There exist two main challenges to detecting the elastic scattering of real photons: (i)~the smallness of the photon-photon scattering cross section, and (ii)~the fact that photons are identical particles such that, e.g., the backscattering of two photons colliding head-on is indistinguishable from the case of no scattering. For two colliding photon beams with finite angular spread, this implies that all the events where both photons are scattered inside the opening angle of the photon beams are undetectable.

Here we show that both the above challenges can be overcome in an asymmetric photon-photon collider setup where a highly collimated broadband gamma-ray beam collides with an extreme ultraviolet (XUV) pulse (see \figref{fig:1}). In fact, a substantial number of scattering events can be triggered by requiring both that photon collisions occur with photon energy in the center-of-momentum (CM) frame $\varepsilon_{\text{CM}} = \sqrt{\varepsilon_\gamma \varepsilon_\omega (1-\cos\vartheta)/2}$ approaching $m_e c^2$, where the elastic photon-photon scattering cross section $\sigma$ is close to its maximum~\cite{Berestetskii-book}, and that at least one of the colliding beams has a large photon density. Here, $\varepsilon_\gamma$ ($\varepsilon_\omega$) is the gamma (XUV) photon energy, $\vartheta$ the angle between the gamma and XUV photon, $m_e$ the electron mass, and $c$ the speed of light in vacuum. In the considered asymmetric setup, the high energy of gamma photons allows us to attain $\varepsilon_{\text{CM}} \lesssim m_e c^2$, while XUV beams provide a relatively large photon number density. Moreover, the high collimation of the gamma beam implies that one or even both of the scattered photons are deflected to angles larger than the beam opening angle, therefore providing an observable of the scattering event (see \figref{fig:1}).

\begin{figure}[tb]
\centering
\includegraphics[width=\linewidth]{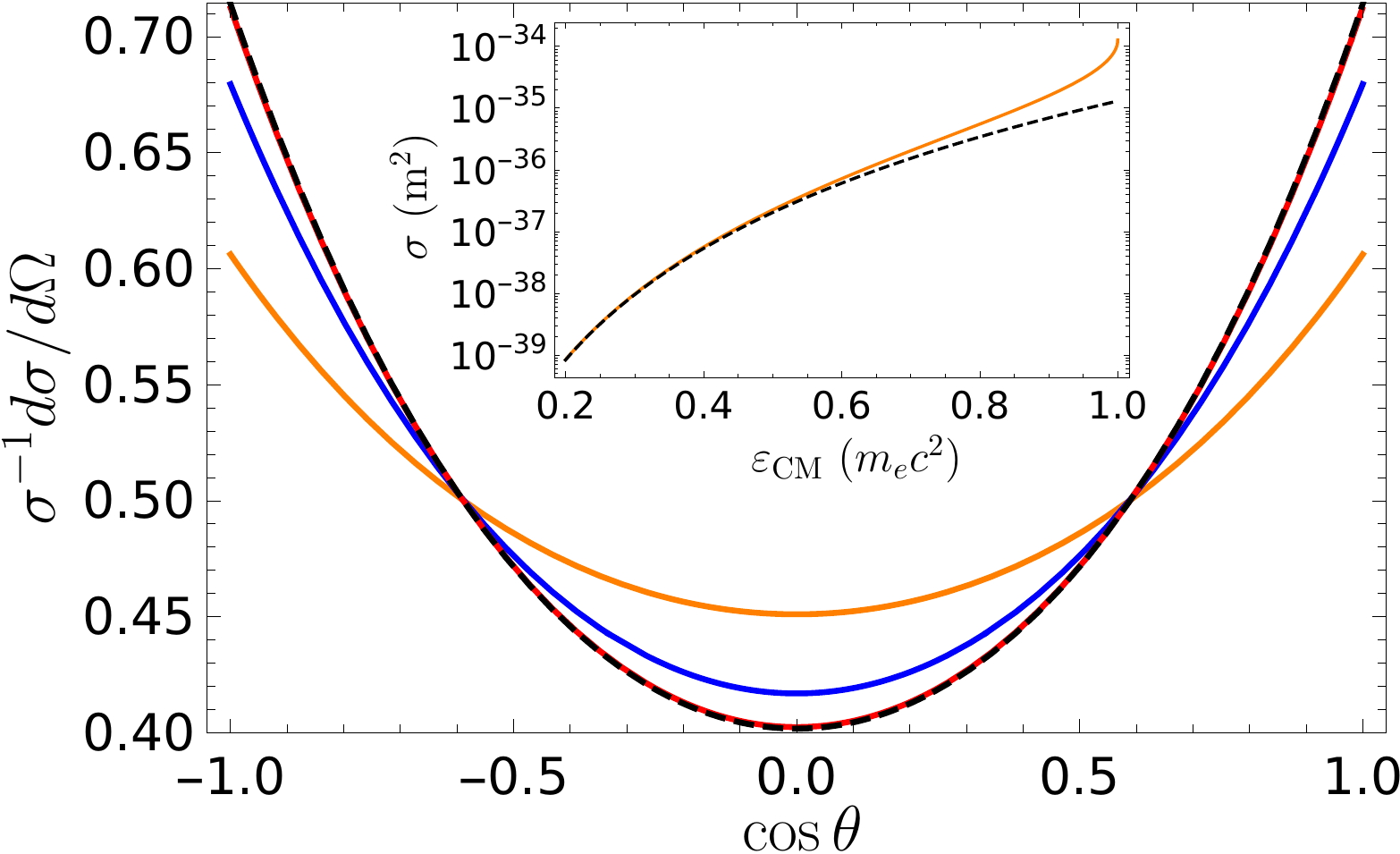}
\caption{Normalized exact differential elastic photon-photon scattering cross section $\sigma^{-1} d\sigma/d\Omega$ in the CM frame for $\varepsilon_{\text{CM}} = 0.9 \, m_e c^2$ (orange line), $\varepsilon_{\text{CM}} = 0.5 \, m_e c^2$ (blue line), $\varepsilon_{\text{CM}} = 0.1 \, m_e c^2$ (red line), and the $\varepsilon_{\text{CM}} \ll m_e c^2$ approximation (black dashed line). The inset plots the exact total elastic photon-photon cross section $\sigma$ (orange line) and its $\varepsilon_{\text{CM}} \ll m_e c^2$ approximation (black dashed line).}
\label{fig:2}
\end{figure}

The elastic photon-photon scattering cross section $\sigma$ reaches its maximum at $\varepsilon_{\text{CM}}$ just above $m_e c^2$ [$\sigma (m_e c^2) \approx 1.3 \times 10^{-34}\text{ m}^2$], drops by 3 orders of magnitude already at $\varepsilon_{\text{CM}} = 0.5 \, m_e c^2$ [$\sigma (0.5 \, m_e c^2) \approx 2.2 \times 10^{-37}\text{ m}^2$], and by a further 4 orders of magnitude at $\varepsilon_{\text{CM}} = 0.1 \, m_e c^2$ [$\sigma (0.1 \, m_e c^2) \approx 1.3 \times 10^{-41}\text{ m}^2$, see \figref{fig:2}]. Due to its strong energy dependence [$\sigma \approx 1.29 \times 10^{-35} (\varepsilon_{\text{CM}}/m_e c^2)^6\text{ m}^2$ for $\varepsilon_{\text{CM}}\lesssim 0.3 \, m_e c^2$ (see \figref{fig:2})], for optical photon beams such as those delivered by the most powerful lasers worldwide~\cite{dansonHPLSE19}, $\sigma$ falls to the tiny value $\sigma(1.55\,\text{eV})\approx 10^{-68}\text{ m}^2$. The smallness of the cross section implies that a large number of photons are required to trigger elastic photon-photon scattering. In fact, the total number of scattering events occurring in the head-on collision of two identical photon beams with focal area $S$ and with $N_\gamma$ photons is $N_\gamma^2 \sigma(\varepsilon_{\text{CM}})/S$. Thus, for 1.55~eV photon energy and $S \approx 10^{-10}\text{ m}^2$, one needs $N_\gamma \approx 10^{29}$ in order to trigger a single scattering event. This huge number of photons corresponds to $2.5 \times 10^{10}\text{ J}$ energy per beam, well above the energy delivered by the world's highest energy facilities such as Laser M\'{e}gajoule~\cite{lmjURL} and the National Ignition Facility~\cite{nifURL}. By contrast, for $0.5 \, m_e c^2$ photons and $S \approx 10^{-10}\text{ m}^2$, one needs $N_\gamma \approx 2.1 \times 10^{13}$ for triggering a single scattering event, which corresponds to an energy of 0.87~J per beam. Thus, a source of large numbers of photons and $\varepsilon_{\text{CM}}\lesssim m_e c^2$ is required.

For $\varepsilon_{\text{CM}}>m_e c^2$, the process of two-real-photon conversion into an electron-positron pair becomes possible~\cite{Berestetskii-book}. Although this is another important QED process that has not yet been observed directly~\cite{pikeNPHO14, thomasNPHO14}, (see Refs.~\cite{burkePRL97} and \cite{adamPRL21} for multiphoton and quasireal photon electron-positron pair creation), its cross section is $\sim\alpha^{-2}$ larger than elastic photon-photon scattering~\cite{Berestetskii-book}, where $\alpha\approx1/137$ is the fine structure constant. Thus, the detection of the much rarer elastic photon-photon process is facilitated in the region $\varepsilon_{\text{CM}} < m_e c^2$, where two-photon $e^-e^+$ pair conversion is kinematically forbidden. In fact, the created $e^-e^+$ pair can yield a relatively large background of photons via Compton backscattering, therefore hiding signs of elastic photon-photon scattering. In addition, the almost complete suppression of background noise associated with Compton scattering of gamma rays with stray particles requires us to operate in extreme high vacuum conditions ($\lesssim10^{-11}\,\text{Pa}$) similar to those of accelerators and storage ring facilities~\cite{redhead:1999fc, xhvURL}. In our setup, background and detector noise can be measured by running the experiment twice: once without the XUV pulse and once for the same conditions but with the XUV pulse (see \figref{fig:1}). Note that detection systems sensitive to single particles in the presence of strong background noise are available \cite{salgadoXXX21} and are at the heart of forthcoming strong-field QED experiments such as the E-320 experiment at the FACET-II beamline~\cite{E320FACET} and the LUXE experiment at DESY~\cite{abramowiczXXX21}.

\begin{figure}[tb]
\centering
\includegraphics[width=\linewidth]{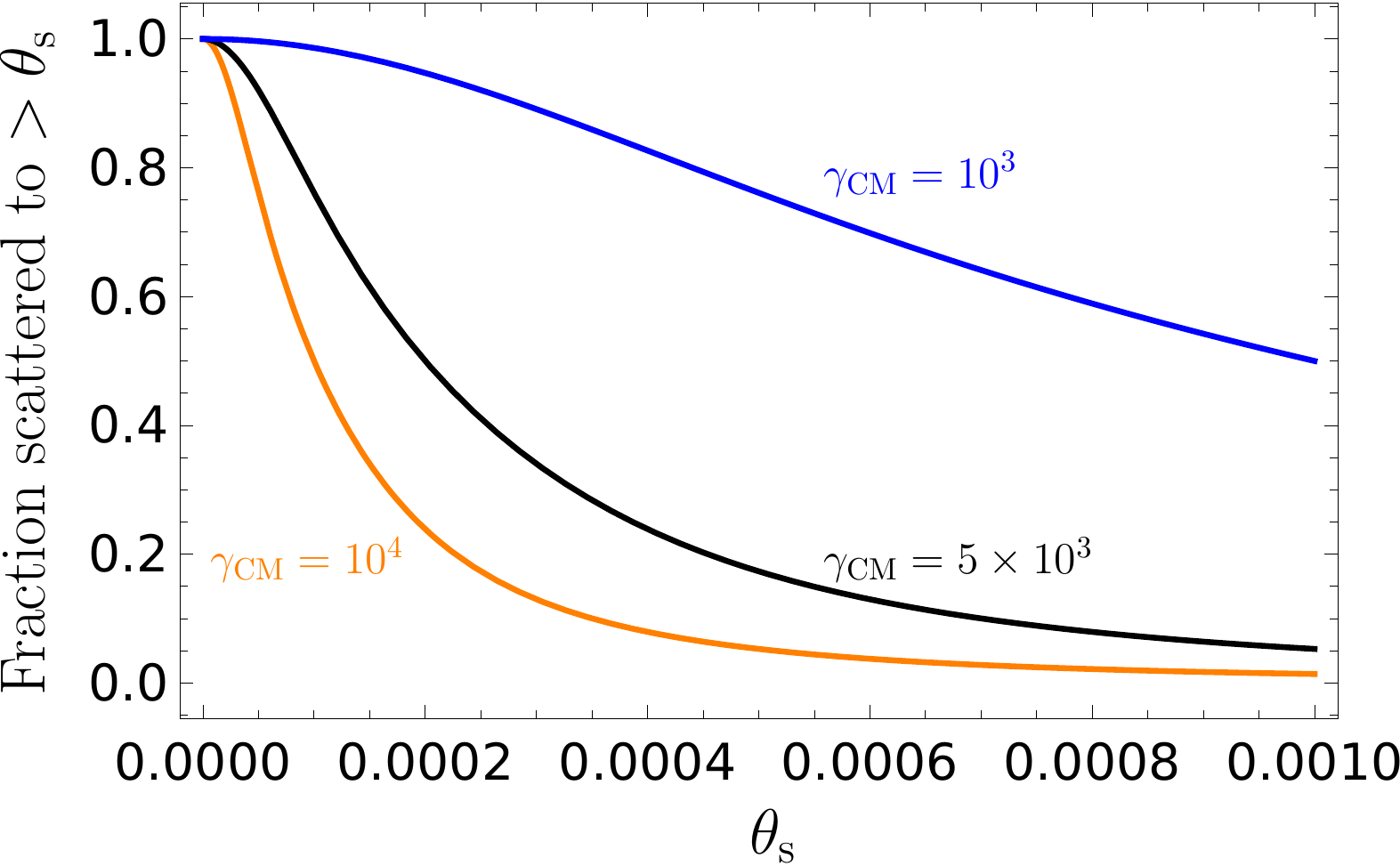}
\caption{Fraction of photons elastically scattered to angles greater than $\theta_s$ from \eqref{eq:dsigma} as a function of $\theta_s$ for $\gamma_\text{CM}=10^{4}$ (orange line), $\gamma_\text{CM}=5 \times 10^{3}$ (black line), and $\gamma_\text{CM}=10^{3}$ (blue line).}
\label{fig:3}
\end{figure}

We start considering the simpler case of a head-on collision of two photons with the same energy, i.e., in the CM frame. Figure~\ref{fig:2} plots the normalized differential elastic photon-photon scattering cross section $\sigma^{-1} d\sigma/d\Omega$ for different values of $\varepsilon_{\text{CM}}$. While the full expression of $d\sigma/d\Omega$ is considerably complex~\cite{karplusPR51, detollisNC65, Berestetskii-book}, \figref{fig:2} shows that the angular distribution of scattered photons weakly depends on $\varepsilon_{\text{CM}}$, such that its much simpler low-energy approximation $\varepsilon_{\text{CM}} \ll m_e c^2$ can be used to infer the key features of the scattered photon distribution. In this case, the unpolarized differential cross section for photons with energy $\varepsilon_\gamma$ and $\varepsilon_\omega$ is
\begin{widetext}
\begin{equation} \label{eq:dsigma}
\frac{d\sigma}{d\Omega} \approx \frac{139 \alpha^2 r_e^2}{64800 \pi^2} \left(\frac{\varepsilon_{\text{CM}} } {m_e c^{2}}\right)^6
\frac{[4 \gamma_{\text{CM}}^2 - 1 - 8 \gamma_{\text{CM}} \sqrt{\gamma_{\text{CM}}^2 - 1} \cos\theta + (4 \gamma_{\text{CM}}^2 - 3) \cos^2\theta]^2} {(\gamma_{\text{CM}} - \sqrt{\gamma_{\text{CM}}^2 -1}\cos\theta)^6},
\end{equation}
\end{widetext}
where $r_e \approx 2.8 \times 10^{-15} \, \text{m}$ is the classical electron radius, $\gamma_\text{CM}:=(1-v^2/c^2)^{-1/2}$ ($\gamma_\text{CM} \approx \sqrt{\varepsilon_\gamma/4 \varepsilon_\omega}$ for a head-on collision with $\varepsilon_\gamma \gg \varepsilon_\omega$), $v$ is the CM-frame velocity in the laboratory frame, and $d\Omega := \sin \theta d \theta d\varphi$ is the solid angle. In \eqref{eq:dsigma} the symmetry factor for identical particles is included, such that $\sigma$ is obtained by integrating $d\sigma/d\Omega$ over the whole solid angle. Notice that the sixth power of $[\gamma_{\text{CM}} - (\gamma_{\text{CM}}^2 -1)^{1/2}\cos\theta]$ appears in the denominator of \eqref{eq:dsigma}, which hints at forward collimation in a cone with $1/\gamma_{\text{CM}}$ opening angle. By analytically integrating \eqref{eq:dsigma} over the whole solid angle except for a region $[0,\theta_s]$ and expanding for $\gamma_{\text{CM}} \gg 1$, one finds an expression that diverges for $\theta_s \rightarrow 0$. Since the integral of \eqref{eq:dsigma} is finite for $\theta_s \rightarrow 0$, this implies that two limits $\gamma_{\text{CM}} \rightarrow \infty$ and $\theta_s \rightarrow 0$ do not commute. This is addressed by setting $\theta_s = r/\gamma_\text{CM}$, where $r$ is an arbitrary constant, and expanding the integral of \eqref{eq:dsigma} for $\gamma_\text{CM} \gg 1$. From the resulting expression, one can note that with excellent approximation 50\% of the photons are scattered outside $1/\gamma_\text{CM}$, while 91\% of the photons are scattered outside $\sqrt{7}/10 \gamma_\text{CM} \approx 0.26/\gamma_\text{CM}$. This dependence on $\gamma_\text{CM}$ is shown by the three curves plotted in \figref{fig:3}, which reports the fraction of photons scattered at angles larger than $\theta_s$ for three different values of $\gamma_\text{CM}$ (see \figref{fig:3}). By setting $\theta_s$ equal to the incident photon beam opening angle, one determines the fraction of detectable scattering events. Thus, $1/\gamma_\text{CM} = \sqrt{4 \varepsilon_\omega/\varepsilon_\gamma}$ should be comparable to or smaller than the photon beam angular aperture for observing photon-photon scattering. Since photon emission by ultrarelativistic electrons in the presence of strong background electromagnetic fields occurs in a cone with $\sim 1/\gamma_e$ opening angle around the electron's momentum~\cite{Baier-book, dipiazzaPRA19}, where $\gamma_e$ is the relativistic factor of the emitting electron, we need to assess when $\gamma_\text{CM} \lesssim \gamma_e$. 


Here we consider gamma-ray generation in the collision of an electron beam with an intense laser pulse with normalized field amplitude $\xi=e E_0/m_e \omega c$, where $E_0$ and $\omega$ are the peak of the electric field and the frequency of the laser pulse, respectively. If $\gamma_e \gg \xi$, then  $\varepsilon_\gamma < \gamma_e m_e c^2$. For a given $\varepsilon_\omega$, both $\varepsilon_{\text{CM}} \approx \sqrt{\varepsilon_\gamma \varepsilon_\omega}$ and  $\gamma_\text{CM} \approx \sqrt{\varepsilon_\gamma/4 \varepsilon_\omega}$ are monotonically increasing functions of $\varepsilon_\gamma$. Thus, $\varepsilon_\gamma \approx \gamma_e m_e c^2$ corresponds to the maximal $\varepsilon_{\text{CM}}$ and $\gamma_\text{CM}$. By choosing $\gamma_e = m_e c^2/\varepsilon_\omega$, both $\varepsilon_{\text{CM}} \lesssim m_e c^2$ and $\gamma_\text{CM} \lesssim \gamma_e/2$ are fulfilled for all photons of the gamma-ray beam. This implies that, in principle, the conditions for observing photon-photon scattering can be always achieved in an asymmetric photon collider. However, in practice, this is limited by the finite electron beam emittance and by the need of high-energy electron beams. Indeed, for optical photons $\varepsilon_\omega=1.55\,\text{eV}$, an ultralow-emittance 168~GeV electron beam would be required, whereas for $\varepsilon_\omega=100\,\text{eV}$ a much more affordable 2.6~GeV electron beam would suffice. Notice that, however, the number of attainable photons per XUV beam decreases with increasing $\varepsilon_\omega$, such that an optimal operational regime is achievable with tens-of-eV photons. 

We consider an electron beam with 0.96~nC charge, Gaussian energy distribution with 13~GeV mean energy, 13~MeV rms energy spread, and 3~mm mrad normalized emittance. The beam spatial distribution has cylindrical symmetry around the propagation axis $x$, and is Gaussian with $20\,\mu$m ($5\,\mu$m) longitudinal (transverse) rms width. The electron beam collides with a 30~TW linearly polarized laser pulse with 30~fs duration FWHM of the intensity and $17^\circ$ electron beam-laser pulse crossing angle (see \figref{fig:1}). The laser pulse intensity is $2.2\times10^{20}\,\text{W/cm}^2$ ($\xi=10$), while its wavelength and waist radius ($1/e^2$ of the maximum intensity) are $0.8\,\mu$m and $3\,\mu$m, respectively. Since the electron beam propagates along $x$ and the laser pulse polarization axis and propagation direction are in the $xy$ plane, large-angle photon emission occurs mainly in the $xy$ plane, with $|\eta_\gamma| := |\arctan{(p_{\gamma,y}/p_{\gamma,x})}|$ of the order of $\xi/\gamma_e$, whereas $|\psi_\gamma| := |\arctan{(p_{\gamma,z}/p_{\gamma,x})}|$ remains of the order of $1/\gamma_e$. Here and in the following, $\bm{p}$ is the photon momentum, and the subscripts $\gamma$ and $s$ denote quantities referring to the gamma-ray beam and to the scattered photons, respectively. Similar electron beam and laser parameters are available at existing facilities such as FACET-II~\cite{yakimenkoPRAB19} and the European XFEL~\cite{exfelURL}, where strong-field QED experiments involving electron beams and laser pulses are planned~\cite{E320FACET, abramowiczXXX21}.

\begin{figure}[tb]
\centering
\includegraphics[width=\linewidth]{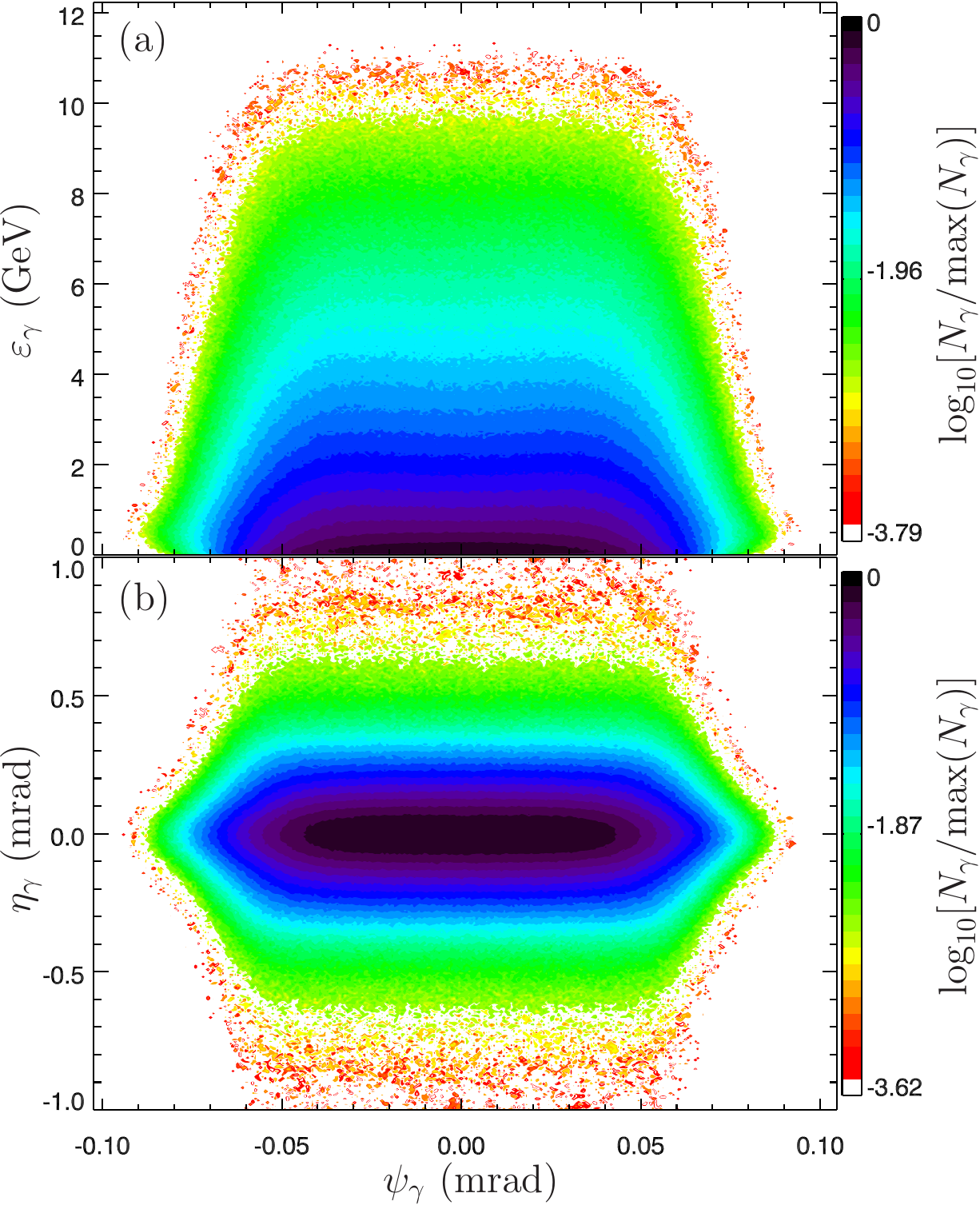}
\caption{Normalized gamma-ray beam distributions after the collimator. (a) $\log_{10}[N_\gamma(\psi_\gamma,\varepsilon_\gamma)/\text{max}(N_\gamma(\psi_\gamma,\varepsilon_\gamma))]$, (b) $\log_{10}[N_\gamma(\psi_\gamma,\eta_\gamma)/\text{max}(N_\gamma(\psi_\gamma,\eta_\gamma))]$. $N_\gamma$ is the photon number, $\varepsilon_\gamma$ the photon energy, $\psi_\gamma :=\arctan{(p_{\gamma,z}/p_{\gamma,x})}$, and $\eta_\gamma:=\arctan{(p_{\gamma,y}/p_{\gamma,x})}$, where $\mathbf{p}_\gamma$ is the photon momentum. Note the different scales for $\psi_\gamma$ and $\eta_\gamma$.}
\label{fig:4}
\end{figure}
\begin{figure}[tb]
\centering
\includegraphics[width=\linewidth]{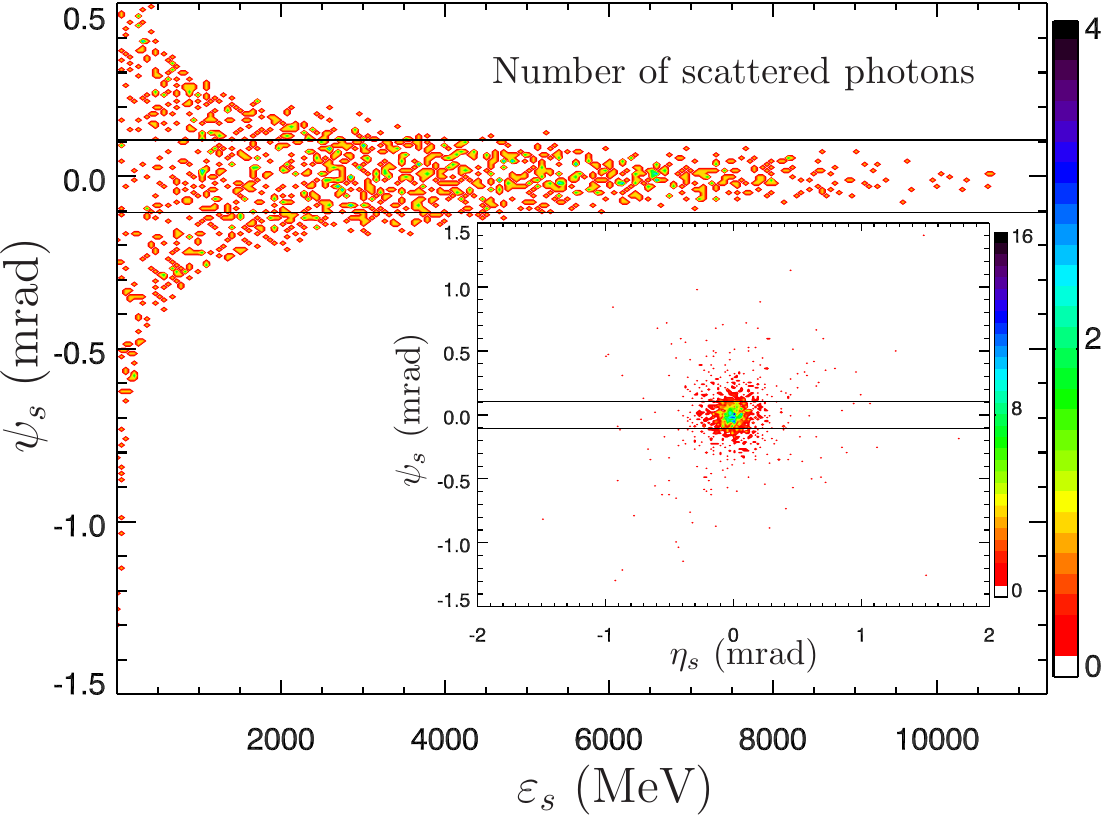}
\caption{Number of elastically scattered photons for one-week operation time as a function of the scattered photon energy $\varepsilon_s$ and of the photon scattering angle $\psi_s:=\arctan{(p_{s,z}/p_{s,x})}$. The inset displays the number of events as a function of $\psi_s$ and of $\eta_s :=\arctan{(p_{s,y}/p_{s,x})}$. Photons scattered in the region outside the two black horizontal lines provide a detectable signal of scattering.}
\label{fig:5}
\end{figure}

After scattering, the electron beam and possible positrons produced in the interaction with the laser are deviated by a 1~T magnetic field directed along $z$ and generated by commercially available permanent magnets with 2~cm length (see \figref{fig:1}). The length of the magnets is chosen such that the 13 GeV electron beam is deflected transversely outside the focal region where the gamma-XUV pulse collision occurs (see below). Subsequently, gamma photons pass through a collimator consisting of two thick parallel plates opaque to gamma rays with 13~cm length along $x$, and $9\,\mu$m interplate distance along $z$. The collimator parameters are chosen such that only well-collimated photons are transmitted while their number remains as large as possible.

The gamma-ray beam distributions after the collimator as a function of $(\psi_\gamma, \varepsilon_\gamma)$ and of $(\psi_\gamma, \eta_\gamma)$ are reported in panels (a) and (b) of \figref{fig:4}, respectively. Details on the methodology for simulating the photon spectrum are reported in Ref.~\cite{dipiazzaPRA19}. As shown in Ref.~\cite{dipiazzaPRA18}, in the collision of an ultrarelativistic electron beam with an intense laser pulse nonlinear effects in the laser field are suppressed for the low-energy emitted photons $\varepsilon_\gamma \ll \gamma_e m_e c^2$. Thus, on one hand the emission of low-energy photons can be described by the standard Klein-Nishina formula even for $\xi \gg 1$. On the other hand, the emission of high-energy photons can be described within the locally constant field approximation~\cite{dipiazzaPRA18, dipiazzaPRA19}. Approximately $1.6\times10^{9}$ photons pass through the collimator, and $\approx2.3\times10^{8}$ of these photons have energy larger than 3.25~GeV; i.e., they reach $\varepsilon_{\text{CM}}>m_e c^2/2$ in the collision with a XUV pulse with 20~eV photon energy (see below).

After passing through the collimator, photons collide with an XUV pulse propagating in the $xz$ plane with a $10^\circ$ crossing angle, and scattered photons with $|\psi_s|>\text{max}(|\psi_\gamma|)\approx 0.1\,\text{mrad}$ are subsequently detected (see \figref{fig:1}). The XUV pulse has 1~mJ energy, 50~fs duration, $8\,\mu$m waist radius, and 20~eV photon energy. Pulses with comparable parameters could be delivered by FELs such as FLASH~\cite{flashURL} or FERMI~\cite{fermiURL}. Alternatively, XUV pulses with tens of eVs of photon energy and $\sim\text{mJ }$ pulse energy can also be generated in high-harmonic generation with xenon~\cite{nayakPRA18}.

Figure~\ref{fig:5} reports the number of elastically scattered photons integrated over a one-week operation time assuming a typical 10~Hz repetition rate (see Refs.~\cite{yakimenkoPRAB19, exfelURL} and Ref.~\cite{rosoEPJWC18} for current electron beam and sub--100~TW-class laser repetition rates, respectively). As a result of a scattering event, the gamma or the XUV photon, or even both photons, can be scattered at relatively large $|\psi_s|$. The average number of elastic two-photon collision events over one day (week) of operation time is approximately 113 (793), which leads to approximately 78 (550) photons scattered in the detectable region, i.e., outside the two black horizontal lines in \figref{fig:5}. Moreover, in approximately 4 (32) of the above-mentioned 113 (793) events over one day (week) both initial photons elastically scatter in the detectable region, therefore providing strong evidence of an elastic photon-photon scattering event. After collision, the energy of the XUV photon is upshifted to the multi-MeV region, with the energy of photons in the detectable region roughly ranging from 5~MeV to 5~GeV (see \figref{fig:5}).
Further increase of the detectable signal is possible by increasing the energy of the electron beam, such that emission occurs in the ``supercritical QED regime'', where electron-to-photon energy conversion efficiency becomes large~\cite{tamburiniPRD21}. Alternatively, the efficient generation of ultradense collimated gamma beams can also be obtained with 10~GeV high-current electron beams colliding with multiple thin foils~\cite{sampathPRL21}.

In order to derive the statistical distribution of scattering events, we start by noting that elastic photon-photon scattering is an extremely rare event where, as confirmed by simulations, in practice each photon scatters at most once in each beam-beam collision. Hence, scattering events occur independently, and the occurrence of one event does not significantly affect the probability that a second event occurs. Under the above conditions, the distribution of scattering events for each gamma-XUV beam collision is known to be Poissonian with mean $\mu \approx n_{\varepsilon_\omega} c \tau \sum_{i=1}^{N_\gamma} \sigma(\varepsilon_{i,\text{CM}}) \ll 1$, where $n_{\varepsilon_\omega}$ is the XUV pulse photon density, $\tau$ is the duration of the XUV beam, and the sum is taken over all photons of the gamma-ray beam. The statistical distribution of scattering events obtained with $N_{b}$ independent gamma-XUV beam collisions, which is the sum of Poisson-distributed random variables, is also Poisson-distributed with mean $M \approx N_b \mu$ and standard deviation $\sqrt{M}$. Thus, in principle, when $M \gtrsim 30$ a possible five standard deviations from the QED prediction become detectable. However, scattering laser and XUV pulse intensity fluctuations naturally alter the number of interacting photons and their energy distribution. Consequently, this changes the expected number of scattering events. Note that relatively low-power laser systems (a 30 TW-class scattering laser is considered here), i.e., with power below 100~TW, can operate with high repetition rate and a precision on the absolute laser intensity below 5\% \cite{abramowiczXXX21}. In order to assess the effect of fluctuations, we have performed further simulations with all parameters fixed but a 5\% intensity fluctuation of both the optical scattering laser and the XUV pulse. Our simulations indicate that a simultaneous 5\% intensity increase or decrease both of the scattering and of the XUV laser can result in approximately a two-standard-deviation change of the number of events. For opposite correlated laser fluctuations, i.e. 5\% intensity increase (decrease) of the scattering laser and 5\% intensity decrease (increase) of the XUV laser, the change of the expected signal reduces to approximately one standard deviation. Note that, however, even a seven-standard-deviation measurement is possible with $M \gtrsim 60$, which is feasible in a single day of operation time (see above).

In conclusion, we have put forward a controllable experimental scheme that allows direct precision measurements of two real photons elastic scattering in vacuum with common lasers at existing facilities. Accurate realistic simulations show that photon-photon scattering can be measured off axis with an asymmetric photon collider within one day of time. Furthermore, this scheme allows us to probe photon-photon scattering across a wide region of sub-MeV CM energies, where any deviation from QED predictions could be evidence of physics beyond the standard model, thereby providing an avenue for direct searches of new physics independent of and complementary to those with electrons and nuclei.

\begin{acknowledgments}
We thank Karen Hatsagortsyan and Christian Ott for useful discussions on the production of intense and energetic XUV beams by using high-harmonic generation in gases. This article comprises parts of the Ph.D. thesis work of Maitreyi Sangal, submitted to the Heidelberg University, Germany.
\end{acknowledgments}

\bibliography{My_Bibliography}

\end{document}